\documentclass[a4paper]{IEEEtran}
\usepackage[english]{babel}
\usepackage{graphicx}
\usepackage{subfig}
\usepackage{booktabs}
\usepackage{amsmath}
\usepackage{systeme}
\usepackage{cases}
\usepackage{pgfplots}

\usepackage{amsthm}
\usepackage{amssymb}
\usepackage{tabularx,tikz}
\usepackage{url}

\usepackage[a4paper,left=1.4cm,right=1.4cm,top=1.9cm,bottom=4.3cm]{geometry}
\makeatletter
\newcommand*{\rom}[1]{\expandafter\@slowromancap\romannumeral #1@}
\makeatother
\begin{document}
\title{Match Made in Heaven: Practical Compressed Sensing and Network Coding for Intelligent Distributed Communication Networks}
\author{\IEEEauthorblockN{Maroua Taghouti,
Anil Kumar Chorppath,
Tobias Waurick and
Frank H.P. Fitzek\\}
\normalsize
\IEEEauthorblockA{Deutsche Telekom Chair of Communication Networks, Technische Universit{\"a}t Dresden, Germany\\}

Email: \{maroua.taghouti$\mid$anil.chorppath$\mid$frank.fitzek\}@tu-dresden.de, twaurick@gmx.net}

\maketitle
\thispagestyle{empty}
\begin{abstract}
Based on the impressive features that network coding and compressed sensing paradigms have separately brought, the idea of bringing them together in practice will result in major improvements and influence in the upcoming 5G networks.
In this context, this paper aims to evaluate the effectiveness of these key techniques in a cluster-based wireless sensor network, in the presence of temporal and spatial correlations. Our goal is to achieve better compression gains by scaling down the total payload carried by applying temporal compression as well as reducing the total number of transmissions in the network using real field network coding. In order to further reduce the number of transmissions, the cluster-heads perform a low complexity spatial pre-coding consisting of sending the packets with a certain probability. Furthermore, we compare our approach with benchmark schemes. As expected, our numerical results run on NS3 simulator show that on overall our scheme dramatically drops the number of transmitted packets in the considered cluster topology with a very high reconstruction SNR.
\end{abstract}
\vspace{-0.3cm}
\IEEEpeerreviewmaketitle
\section{Introduction}

According to Ericson, around 28 billion devices are predicted to be connected to the network world wide by 2021 and the ubiquitous massive Internet of Things (IoT) is an integral part of the evolution towards the 5G standard \cite{EricssonIoT}. This will result in massive data procreation, where just the mobile data traffic will increase 7-fold by 2021 \cite{CiscoVNI} that must be handled by the fundamental techniques considered for 5G. We consider the use case of industrial automation with massive deployment of 500 to 2000 sensors in a smart factory. The sensors are clustered and are connected to the cluster heads through an IoT Gateway. The cluster heads form a mesh network between each other to deliver the sink packets for reconstructing sensor data. The traffic going from the sensors through the IoT gateway and the overall data transmissions in the mesh network needs to be limited. The goal is to extend the battery life of the sensors, IoT gateways and cluster heads as well as to reduce latency for some critical applications.

Network coding \cite{ahlswede2000network} \cite{ho2006random} is a disruptive paradigm for reliable data dissemination as it performs similarly to an Forward Error Correction (FEC) but with more robustness and flexibility. However, it is considered as an ``all-or-nothing" code, which means that the receiver would need at least as many coded packets as the number of the original ones in order to decode. It is yet possible to recover a partial fraction with a low probability \cite{claridge2017probability}. This becomes critical for delay sensitive transmissions such as multimedia streaming, since coded packets would require usually some delays before being decoded. Additionally, it is possible that the whole decoding process fails even when one single coded packet is not recieved. On the other hand, compressed sensing \cite{candes2006stable} \cite{donoho2006compressed} advocates for finding exact solutions to sparse under-determined systems contrarily to network coding, using reconstruction algorithms.

Both techniques have been widely investigated in wireless networks, but mostly without investigating their fundamental interplay. Therefore, bringing them together in practice will have a large impact on the upcoming 5G networks, IoT and Machine-to-Machine communications, all of which are meant to serve a massive amount of devices and traffic. Furthermore, this combination is expected to be critical in the sense that it will enable millisecond response time in large-scale sensor networks. 

Taking into consideration the fact that Wireless Sensor Networks (WSN) have limited battery life, connected through unreliable wireless links and that the data harvested are usually correlated, we aim to oppose these constraints by proposing an intelligent JOint COMpressed sensing and network COding (JoComCo) framework for cluster-based topologies that exploits the inter and intra sensors correlations to reduce the payload sizes and the number of transmitted packets within the network. In a nutshell, the sensors inside a cluster perform temporal compression on their acquired data before conveying them to the cluster after an election stage called spatial pre-coding. Later, the cluster heads perform real field network coding on their data and the incoming ones from other clusters in order to improve their delivery. The spatial pre-coding and real field compressed sensing operations form the elements of sensing matrix in the compressed sensing operation. Joint decoding at the sink is ensured by a greedy compressed sensing reconstruction algorithm.
Moreover, we investigate the jointly optimal values of transmission probability and network coding spatial compression ratio for the topology. Our design improves the robustness against packet losses along the paths and promotes the sparse reconstruction process at the sink nodes in terms of number of transmissions, complexity, delay as well as online capabilities.

Our results obtained using the NS3 simulator and Orthogonal Matching Pursuit (OMP) algorithm show that the total number of packet transmissions can be reduced considerably by the joint scheme especially when the data correlation is high. The results also show the graceful improvement in the error correction using the proposed scheme with increasing redundancy added.  

The rest of the paper is structured as follows. Section \rom{2} proposes the motivation behind such a joint design and highlights the related works. Section \rom{3} concisely introduces the system model and all the steps of the proposed scheme design. Section \rom{4} explains the reconstruction and the theoretical gains. Section \rom{5} discusses the implementation using NS-3 simulator and explains the numerical results. 
\section{Motivation and Related Works}
This section presents some incentives for bringing network coding and compressed sensing together. It highlights the accompanying potentials for improving current and future networks with massive devices. Moreover, it emphasizes the state-of-the-art joint schemes as well as what they lack.
\subsection{Cross Potential}
When performing only network coding in large-scale Wireless Sensor Networks (WSNs), we benefit from interesting features that enhance the network performance in general, for instance error correction, exploiting path diversity and the broadcast nature of the wireless sensors. However, the ``all-or-nothing" critical drawback will refrain network coding from drastically cutting off the number of transmissions in spite of the spatial correlations. Distributed compressed sensing \cite{baron2006distributed} used among intra and inter nodes is capable of overcoming the over-determined system constraint but fails to keep such a reduced number of transmissions in the presence of errors. On the other hand, the idea of a joint agnostic compressed sensing and network coding scheme has the potential of being more efficient in the sense that together they reduce the payload and the number of packets transmitted even in lossy environments. In this case, the decoding cannot be accomplished faster because it requires multiple independent decoding steps consisting of inverting the previous ones performed during the encoding.
Therefore, in order to embrace the effectiveness of a joint combination approach, both techniques should operate in the same field type, i.e. using either finite fields or the real field. In such a manner, joint decoding is possible. Subsequently, it is expected to have lower computational complexity as well as lower energy consumption.
Compressed sensing is based on random projections and its main reconstruction algorithms rely on $\ell_p$ norms, $p \leq 1$, which are not mathematically logical to be computed in the finite field unlike what is claimed in \cite{kwon2014compressed}. It is accordingly reasonable to propose a scheme that operates exclusively on the real field. Network coding in the real field was first introduced in \cite{katti2007real}, and later extended to the complex field \cite{dey2008real}. The coding coefficients are i.i.d drawn from a random distribution such as the Gaussian or the Bernoulli/Rademacher distributions as they prove to be compatible with compressed sensing properties with a high probability while preserving some features of random linear network coding.
\vspace{-0.15cm}
\subsection{Joint Schemes: State-of-the-art}
To the best of our knowledge, there are only few works that dealt with combining both approaches beforehand. Some general approaches based on quantization before applying network coding were presented in \cite{nabaee2012quantized} \cite{nabaee2014quantized}, as well as specific applications that cover for example distributed storage \cite{yang2013energy} and spectrum sensing \cite{dehghan2012network}.
The ``Netcompress" scheme was proposed in \cite{Nguyen2010NetCompress}, where the sensor data is compressed in real field at the relay nodes, similarly to the recoding in traditional random linear network coding. The reconstruction error is obtained with the number of measurements. The header size explosion problem is dealt by using a sparse variation of Bernoulli coefficients along with discards for some cases. However, their reconstruction gain is limited to the one obtained with the standard compressed sensing, i.e.payload reductions. In \cite{feizi2011power}, Feizi et al. suggested a robust scheme that is aware of the cross potential in coupling both techniques. They show theoretical gains that surpass the aforementioned scheme. However, no implementation and evaluation of the scheme with joint spatial pre-coding at source and network coding were carried out. Spatial and temporal correlations of data in sensor networks with cluster-based topology is exploited by joint network coding and compressed sensing in \cite{chen2016compressive}. The scheme's evaluation is based on temporal compression ratio and the relation of network coding operation to spatial sparsity is not evaluated. Additionally, there is no evaluation based on the number of packet transmissions in the network compared to the cases with only network coding or temporal compressed sensing. 
\vspace{-0.3cm}
\section{Joint Compress and Code Scheme (JoComCo)}
\label{sec: model}
In a nutshell, our scheme is different is the sense that 
we carry out spatial pre-coding for energy saving of the sensors in the setting of \cite{chen2016compressive}, so that some sensors do not transmit to the cluster head. In contrast to \cite{chen2016compressive} we consider the case with measurement noise. We also consider the effect of link errors and perform error corrections thanks to network coding. Intuitively, the higher the number of encoded packets and recombinations, the more the redundancy to compensate for the packet losses in the links.
\subsection{System Model: Cluster-based Topology}
\vspace{-0.15cm}
\begin{figure}[h!]
    \centering
    \includegraphics[width=6.5cm]{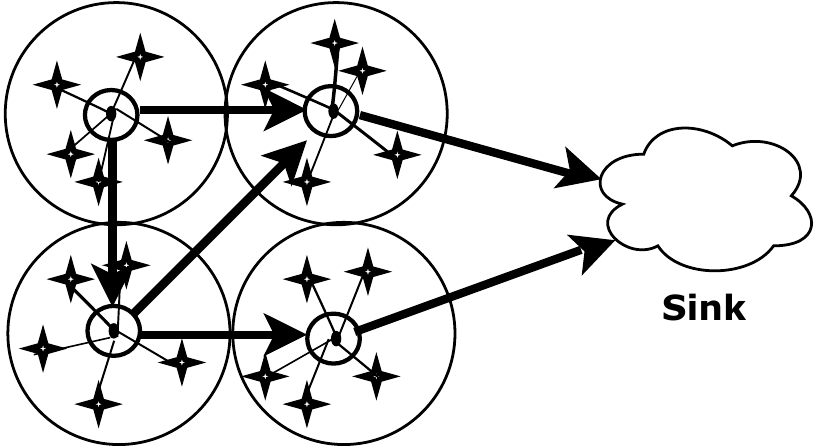}
    \caption{Clustered Topology comprising $C=4$ clusters and $N=512$ sensors in total including the cluster head. Arrows represent the flows between cluster heads, as well as the sink (a combination of tree and diamond topology).}
    \label{Topology}
\end{figure}
Without loss of generality, we define a cluster topology composed of $C$ clusters, $C \geq 1$, each of which has $N_i$ sensing nodes, $i \in \mathcal{C}=\{1, \cdots , C\}$, resulting in $N$ sensors in total for this topology, as in Figure \ref{Topology}.
The sensor nodes take simultaneously and periodically measurements that they store in batches of size $n$ each before any processing is done. All nodes acquire one reading per time slot. We assume that these readings can be mapped to lower dimensional spaces.
 Let $x_{ij}= [x_{ij,1}, \cdots , x_{ij,n}]$ be the vector of $n$ readings, $n \in \mathbb{N}^*$, of the node $j$ in cluster $i$.

We assume that each $x_{ij}$ can be mapped to lower dimensional spaces and follows the Joint Sparsity Model JSM-2 for temporal and spatial correlated signals, known as a common sparse supports model, i.e. the signals have the same sparsity pattern but with different non-zero elements \cite{baron2006distributed}. Basically the signals can be seen as:
\begin{equation}
\hat{x}_{ij}=x_{ij} + w_{ij} ,
\end{equation}
\begin{equation}
\hat{x}_{ij}=\mathbf{\Psi} \theta_{ij} + w_{ij} , \quad j \in \{1,\cdots, n\},
\end{equation}
where $\mathbf{\Psi} \in \mathbb{R}^{n \times n}$ is the temporal sparse basis that is assumed to be the same for all nodes, $\|\theta_{ij}\|_0=k$ and $w_{ij}$ is the additive white Gaussian noise. Eventually, all measurements in the cluster $i$ are:
\begin{equation}
\mathbf{x}_i^T=[x_{i1} \quad x_{i2} \quad \cdots \quad x_{iN_i}]^T.
\end{equation}
\vspace{-0.3cm}
\subsection{Temporal Pre-coding}
As the sensors are assumed to be capable of sampling their data, each sensor $i$ in a cluster $j$ compresses its original readings $x_{ij}$ of size $n$ into an $m$ dimensional vector, where $m \ll n$ as the compressed sensing theory stipulates. This can be locally performed using a pseudo-random number generator seeded with its identifier, here $j \in  \{1, \cdots, n\}$. Let $y_{ij} \in \mathbb{R}^{m \times 1}$ be the resulting compressed signal obtained as follows:
\begin{equation}
y_{ij}= \mathbf{\Phi} x_{ij},
\end{equation}
where $\mathbf{\Phi} \in \mathbb{R}^{m\times n}$ is the transform matrix with random entries used for temporal compression per sensor which satisfyies the Restricted Eigenvalue (RE) condition \cite{raskutti2010restricted}. We assume that all the sensors are observing the same phenomenon or process. Therefore the temporal sparse basis can be assumed to be the same for each sensor. Note that $\mathbf{\Phi}$ is only required to be known by the sink. Finally, let $\mathbf{Y_i} \in \mathbb{R}^{m\times N_i}$ be the matrix of all temporally compressed vectors: 
\begin{equation}
\mathbf{Y}_i= [y_{i1} \quad y_{i2} \cdots y_{in}]^T.
\end{equation}
\vspace{-0.3cm}
\subsection{Spatial Pre-coding}
\vspace{-0.1cm}
The spatial pre-coding step consists of transmitting the compressed data with a probability $p_{i}$ because they are geographically close to each other, i.e. there is a high probability of being spatially correlated. Literally, every sensor decides its transmission probability on its own in a distributed way and the resulting packets are obtained as follows:
\begin{equation}
\mathbf{Z}_i= \mathbf{B}_i \mathbf{Y}_i
\end{equation}
where $\mathbf{B_i} \in \mathbb{R}^{a_i \times N_i}$ is the pre-coding matrix and $a_i$ depends on the transmission probability $p_{i}$. The overall pre-coding matrix $\mathbf{B}$ can be written as a diagonal matrix with $\mathbf{B}_i \, \ \forall i \leq N$ considering $\mathbf{B}_i$ as the diagonal sub-matrices.

If the transformation matrix $\mathbf{\Phi}$ satisfies the RE property, \cite{feizi2011power} suggests choosing binomial distributed elements.
For the spatial pre-coding, we introduce a modification: 
\begin{equation}\label{eq:3-feiziBi}
p(b_{i}=1) = 1-p(b_{i}=0) = \frac{\mu (l-1)}{N-1}, \ \forall i \in \mathcal{C}
\end{equation}
where $\mu$ is the scaling factor to ensure that $l-1$ or more packets are received with high probability. We select $\mu$, where $(\mu -1) (l- 1) = 2 \sigma$
so that the probability of having more than or equal to $(l-1)$ transmissions is approximately 0.9978, where $\sigma$ is the standard deviation of the aforementioned binomial distribution.

The transmission probability of the sensors is adjusted according to the required reconstruction SNR at the sink. However, this will create a varying number of received packets at each cluster heads. Therefore, combining spatial pre-coding with network coding at cluster heads, constant number of enough output packets can be ensured from the cluster heads to the sink.
\subsection{Network Coding}
\paragraph{Intra cluster}
Now that the selected data $\mathbf{Z}_i$ in each cluster, ($\forall \ i \in \mathcal{C}$), are transmitted to the cluster head (including its own data with a certain probability), the first stage of network coding is at the cluster heads in the network. Let $\mathbf{\Omega}_i$ is the sensing matrix to capture the network coding at the cluster head $i$. 
\begin{equation}
\mathbf{H}_i = \mathbf{\Omega}_i \mathbf{Z}_i,
\end{equation}
where $\mathbf{\Omega}_i \in \mathbb{R}^{l_i \times a_i}$, $l_i \leq a_i$,  $l_i$ is the number of outgoing packets from cluster head $i$. $\omega_i$ is a matrix with coefficients drawn i.i.d from the Bernoulli distribution as it is expected to have less quantization loss.

Let $\mathbf{\Omega}$ be the overall network coding matrix, which records all current network coding operations inside each cluster, i.e. it is designed in such a manner that $\mathbf{\Omega}_i$, $\forall i \in \mathcal{C}$ are the diagonal sub-matrices as follows:
\begin{equation}
\mathbf{\Omega} = \begin{bmatrix}
	\mathbf{\Omega}_{1} & 0              & \cdots & 0              \\
	0              & \mathbf{\Omega}_{2} & \cdots & 0              \\
	\vdots         & 0              & \ddots & 0              \\
	0              & \cdots         & 0      & \mathbf{\Omega}_{C}
\end{bmatrix} 
\end{equation} 
\paragraph{Inter cluster}
The next stage of network coding is the re-combinations at the intermediate cluster heads where the incoming packets from other clusters are coded along with packets of the coding cluster. Here, the coefficients are chosen at random from a Gaussian distribution because the Bernoulli distribution property for reconstruction will simply not hold if we use it more than one time.
Let $\mathbf{\Omega}^r$ be the sensing matrix for this. Therefore, the entire network coding operations can be captured by $\mathbf{\Omega} \mathbf{ \Omega}^r$. They are designed such that they jointly conserve the RE property. Note that random encoding matrices drawn from such distributions are full-rank with a high probability for a sufficiently large $a_i$. Furthermore, the network coding operations at the cluster heads help dropping the total number of redundant transmissions in the network thanks to the effectiveness of network coding in the presence of lossy links. As such, we introduce the redundancy at the sender of each link proportionally to the packet loss probability. 

\textbf{Remark:} We observe that a normalization of the Gaussian coefficients is required for the network coding operations so that the cascading operations do not affect the coefficients and the matrix remains Gaussian. Normalization by $\frac{1}{\sqrt{q}}$ is required at a node where $q$ is the number of outgoing packets from that node.
\section{Joint Reconstruction}
\subsection{Reconstruction}
The types of compressed sensing and network coding presented in this paper preserve the features required for efficient compressed sensing decoding using greedy reconstruction algorithms, namely OMP \cite{needell2009uniform}.
Only one decoding type that can handle all previous steps performed for encoding. It consists of two steps, namely spatial decoding and temporal decoding. Let $Y= [y_{i1}, \cdots, y_{iN}]^T$ be the matrix of all temporally compressed vectors in the whole network, $\mathbf{Y} \in \mathbb{R}^{m\times N}$. The received values at the sink from the $l \ll N$ packets are 
\begin{equation}
\mathbf{U} = \mathbf{\Omega^r} \mathbf{\Omega} \mathbf{B} Y
\end{equation}
where $\mathbf{U} \in \mathbb{R}^{l\times m}$.  Let $\widetilde{Y}\in \mathbb{R}^{N\times m}$ be the solution to the spatial compressed sensing reconstruction problem corresponding to each sensor \cite{candes2006stable}. Several convex or greedy algorithms are available for reconstruction \cite{needell2009uniform}, \cite{needell2009cosamp}. A total of $m$ spatial reconstruction is carried out to obtain the $Nm$ values using the $l$ packets each with $m$ values.

In the next stage, a total of $N$ temporal compressed sensing reconstruction is carried out to obtain the $n$ values of each sensors. Let $\widetilde{y}_{ij}$ be one element of the spatially reconstructed $\mathbf{\widetilde{Y}}$. Then this stage recovery determines $\widetilde{x}_{ij}$ as the sparse solution to
\begin{equation}
\widetilde{y}_{ij}= \mathbf{\Phi} \widetilde{x}_{ij}.
\end{equation}
As for the Reconstruction SNR (RSNR), it is defined as
$$ \text{RSNR} = 20 \log_{10} (\frac{||x||_{l_2}}{||x - \widetilde{x}||_{l_2}}) dB$$
where $\widetilde{x}$ is the reconstructed vector and reconstruction error is $ ||x - \widetilde{x}||^2_{l_2}.$. We use this parameter for the evaluation of different schemes in Section \ref{sec: Implementation}.

In the cluster topology, there are $N_i$ sensors under each cluster head node. If compressed sensing and network coding are carried out separately, the head nodes send all the $l$ packets they receive from the sensors to the sink. By doing them jointly, i.e. compressing within the network, the head nodes need to transmit only few combinations of the $l$ values to the sink to reconstruct the $Nn$ values. When spatial pre-coding is used, the number of source packets received at the cluster heads should be greater than or equal to the required number of output packets from the cluster heads.

 



\subsection{Theoretical Compression Gain}
As a consequence to the temporal compression, every packet's payload to transmit $y_{ij}$ is reduced by a factor of 
\begin{equation}
\upsilon_T = \frac{m}{n},
\end{equation}
which is quite small since $m \ll n$. From a practical point of view, this would reduce the probability of having erasures in the packets. The overall compression gain is defined as $\upsilon_S \upsilon_T$ , where $\upsilon_S $ is the gain in number of total transmissions in the network due to spatial compression and $\upsilon_T$ is the gain in payload size of packets due to temporal compression. Let us consider the single cluster case first. With only spatial compression, the compression gain is $\frac{N-1+l}{2N-1}$, where $l$ is the number of packets sent from cluster head to the sink for the reconstruction. The value of $l$ is selected to ensure that the required RSNR is achieved with high probability. 

The compression gain achieved by joint CS and NC scheme is:
\begin{equation}
g_{JoComCo}^{(1)}=\frac{N-1+l}{2N-1}\frac{m}{n}.
\end{equation}
Whereas with pre-coding in addition it becomes:
\begin{equation}
g_{JoComCo, precode}^{(1)}=\frac{\mu l+l}{2N-1}\frac{m}{n}
\end{equation}

\subsection{Error Correction}
When there are packet losses in the links, the network coding operations at the cluster heads help in reducing the total number of redundant transmissions in the network. With the joint network coding and compressed sensing scheme, the exact packets are not required to be retransmitted in case of packet loss as in ARQ and only more randomly generated coded packets are required. This will improve also the overall latency of a block of packets compared to the ARQ scheme with retransmissions.  The redundant coded packets will compensate for the losses and improve the probability of achieving the minimum RSNR required at the sink for reconstruction, even with redundancy less than the value proportional to the packet loss probability. With higher redundancy on coded packets added to the sender of each link the better will be the probability of achieving the minimum RSNR. The evaluation of the error correction performance of the JoComCo scheme is given in the next section.
\section{Implementation and Results}
\label{sec: Implementation}
In this section, we evaluate the proposed JoComCo scheme using NS3 simulator along with the KL1p \cite{kl1p} library for compressed sensing. Since KL1p does not explicitly calculate the sensing matrix, but rather recreates its multiplication each time, we improved upon this by providing the sensing matrix and our simulations run at least 50 times faster. We implemented the clustered topology given in Figure \ref{Topology}, which contains 4 clusters and multiple hops. The number of source nodes per cluster is $N_i=128$. Thus, the total number of packets to be eventually transmitted without compression and coding would be $N=512$. Note also that we fixed $n=512$ and $m=64$. We set the initial SNR with respect to the Gaussian measurement noise at each sensor to 150 dB.

We consider four schemes for performance comparison, namely (i) ``Only NC", where only network coding is applied in the cluster heads, (ii) ``Only temp. CS", which is the case where only temporal compression is applied by the sensors, (iii) ``JoComCo w/o prec", which is our proposed scheme without the pre-coding step, (iv) ``JoComCo w prec", which is the enhanced one based on pre-coding.

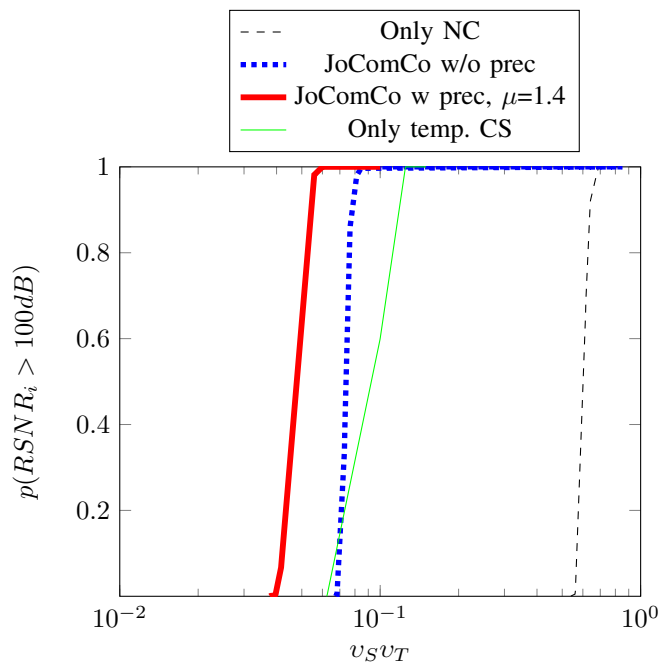
\begin{figure}
\begin{tikzpicture}
  \begin{axis}[ 
    ylabel={$p(RSNR_i > 100dB)$},
    xlabel={$\upsilon_S \upsilon_T$},
		ymin=0, ymax=1,
    xmin=0.01, xmax=1,
		xmode=log,
    xtick={0.01, 0.1, 1},
    ytick={ 0.2, 0.4, 0.6, 0.8, 1},
		legend style={at={(0.21,1.2)},anchor=west}
  ] 
	
	\addplot[
    color=black,dashed
        ] coordinates {
(0.540117416829746,	0) (0.542074363992172,	0.001)
(0.559686888454012,	0.005015625) (0.561643835616438,	0.012) (0.614481409001957,	0.655078125) (0.616438356164384,	0.69005859375) (0.639921722113503,	0.91101171875)(0.64187866927593,	0.92201171875) (0.682974559686888,	0.99100390625) (0.684931506849315,	0.993) (0.69,	1) (0.85,	1)
};
\addlegendentry{Only NC}

	\addplot+[
    line width=2pt, color= blue, mark=none, dotted
    ] coordinates {
(0.0675146771037182,	0)(0.0677592954990215,	0.002)
(0.0680039138943248,	0.003)
(0.0682485322896282,	0.00300390625)(0.0728962818003914,	0.33789453125) (0.0731409001956947,	0.38790234375) (0.0765655577299413,	0.85365625)(0.0768101761252446,	0.86464453125) (0.0812133072407045,	0.97854296875)(0.0814579256360078,	0.98053515625) (0.0841487279843444,	0.9955078125) (0.0843933463796478,	0.9965078125) (0.85,	1) (0.1,	1)
};
\addlegendentry{JoComCo w/o prec}

\addplot+[
    line width=2.2pt, color=red, mark=none
    ] coordinates {
(0.0377208904109589,	0)(0.0379655088062622,	0.001)(0.0394332191780822,	0.002) (0.0396778375733855,	0.00500390625)(0.0416347847358121,	0.0670234375) (0.0418794031311155,	0.085953125)  (0.0558226516634051,	0.98151953125) (0.0560672700587084,	0.98251171875)(0.06,	1) (0.1,	1)
};
\addlegendentry{JoComCo w prec, $\mu\text{=1.4}$}

\addplot[
    color=green
    ] coordinates {
(0.0623776908023483,	0) (0.0626223091976517,	0.00390541076660156) (0.0998043052837573,	0.597527847290039) (0.100048923679061,	0.601433258056641) (0.124755381604697,	0.995879745483398) (0.125,	0.99978515625) (0.13,	1)(0.15,	1)
};
\addlegendentry{Only temp. CS}

  \end{axis}
\end{tikzpicture}
\caption{CDF of the compression gain at the sink.}
\label{pictureCDF}
\end{figure}

Figure \ref{pictureCDF} displays the reconstruction probability of the schemes, with respect to a certain Reconstruction SNR, as a function of the compression gain $\upsilon_S \upsilon_T$, which gives the total spatial and temporal compression achieved by a scheme. Note that here we consider a single cluster topology for an in-depth understanding of the compression and coding inside one cluster. As expected, the combination of compressed sensing and network coding outperforms the case where only network coding is used, even though this presents a slight improvement compared to conventional routing in WSN networks consisting of sending at best a ratio of $0.69$. As for the ``JoComCo w/o prec", it dramatically drops this ratio to $0.084$, and to $0.06$ when pre-coding is performed. This can result in important improvements especially in the sensors' battery lives and in alleviating the network traffics. Additionally, relying only on the temporal compression of the readings will not guarantee such a result as its compression gain is $0.125$ with a high probability for such a topology. 

\begin{figure}
\begin{tikzpicture}
  \begin{axis}[ 
  grid = major,
    xlabel={$ ck$},
    ylabel={$\upsilon_S \upsilon_T$},
		ymin=0, ymax=1,
    xmin=0, xmax=0.12,
    xtick={0, 0.03, 0.06, 0.09, 0.12},
    ytick={0, 0.2, 0.4, 0.6, 0.8, 1},
		legend style={at={(0.03,0.538)},anchor=west}
  ] 

\addplot[
    color=red,
    mark=+,
    ] coordinates {
(0.0138106793200498,0.785234899328859)(0.0276213586400995,0.785234899328859)(0.0414320379601493,0.785234899328859)(0.055242717280199,0.785234899328859)(0.0690533966002488,0.785234899328859)(0.0828640759202985,0.785234899328859)(0.0966747552403483,0.785234899328859)(0.110485434560398,0.785234899328859)
};
\addlegendentry{Only NC}
\addplot[
    color=blue,
    mark=square,
    ] coordinates {
(0.0138106793200498,0.0920197147651007)(0.0276213586400995,0.134962248322148)(0.0414320379601493,0.177904781879195)(0.055242717280199,0.220847315436242)(0.0690533966002488,0.263789848993289)(0.0828640759202985,0.306732382550336)(0.0966747552403483,0.349674916107383) (0.110485434560398,0.39261744966443)
};
\addlegendentry{Separate NC, temp. CS}
\addplot[
    color=black,
    mark=*,
    ] coordinates {
(0.0138106793200498,0.0479760906040268) (0.0276213586400995,0.0757480425055928) (0.0414320379601493,0.106945609619687) (0.055242717280199,0.141568791946309) (0.0690533966002488,0.179617589485459 ) (0.0828640759202985,0.221092002237136) (0.0966747552403483,0.265992030201342)(0.110485434560398,0.314317673378076)
};
\addlegendentry{JoComCo w/o prec}

\addplot[
    color=green,
    mark=diamond,
    ] coordinates {
(0.0138106793200498,0.0243806365352349)(0.0276213586400995,0.0448539918903803)(0.0414320379601493,0.0711069193931767)(0.055242717280199,0.103122640520134)(0.0690533966002488,0.140905113954139)(0.0828640759202985,0.184567472385347)(0.0966747552403483,0.2339524433724)(0.110485434560398,0.289151286353468)
};
\addlegendentry{$\text{JoComCo w prec,}\mu\text{=1.2}$}
  \end{axis}
\end{tikzpicture}
\caption{Gain in number of packets with spatial data correlation.}
\label{picturesparse}
\end{figure}
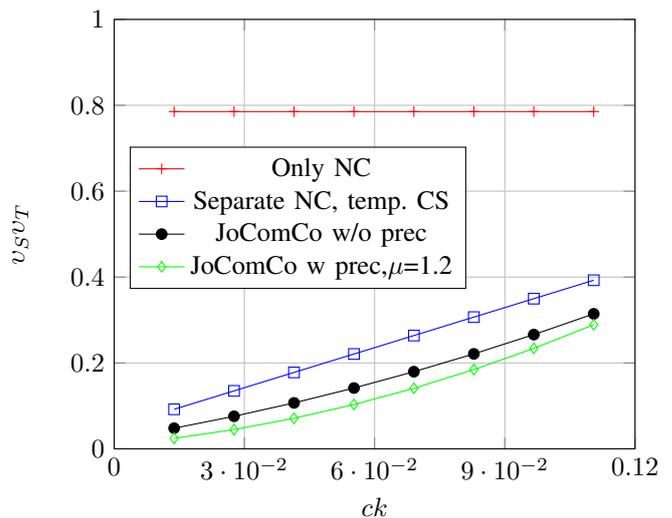

Figure \ref{picturesparse} depicts the average compression gain $\upsilon_S \upsilon_T$ as a function of the normalized sparsity expressed as $ ck$, where $c=\sqrt{1/N^2+1/n^2}=\frac{\sqrt{2}}{512}$, for the different schemes when adopting the topology presented in Figure \ref{Topology}. It is clear that the sparsity does not impact the number of packets needed for decoding when using only network coding. Therefore, we would still require a fraction of around $0.8$ in average to decode. However for the remaining schemes, the sparsity factor would play an important role in reducing it as $k$ grows. For example, if we agnostically combine the temporal compression and the network coding schemes, for a normalized sparsity of $0.013$, the gain would be $0.09$ in average. As for our advocated scheme with pre-coding, we are even closer to $0.024$ for the same parameters, which is a noticeable enhancement compared to state-of-the-art scheme, including ``JoComCo w/o prec". We conclude that the performance of such a scheme is highly correlated with the sparsity factor and the ``JoComCo w prec" ensures a high reduction in terms of the coded packets to send in such a topology compared to the case where only network coding is performed.

\begin{figure}
\begin{tikzpicture}
  \begin{axis}[ 
  grid = major,
    xlabel={Normalized redundancy, $\rho$},
    ylabel={$p(RSNR_i > RSNR_{min})$},
		ymin=0, ymax=1,
    xmin=1, xmax=1.7,
    xtick={1.1, 1.2, 1.3, 1.4, 1.5, 1.6, 1.7},
    ytick={ 0.2, 0.4, 0.6, 0.8, 1},
		legend style={at={(0.3,0.3)},anchor=west}
  ] 
\addplot[
    color=red,
    mark=+,
    ] coordinates {
(1.17,0.965)(1.25,	0.973) (1.33,	0.981) (1.428,	0.9856) (1.53,	0.991) (1.53,	0.998) (1.67,	0.999)};
\addlegendentry{$RSNR_{min}=20dB$}
\addplot[
    color=blue,
    mark=*,
    ] coordinates {
(1.17,0.48)(1.25,	0.49) (1.33,	0.56) (1.428,	0.7) (1.53,	0.78) (1.53,	0.78) (1.67,	0.83)};
\addlegendentry{$RSNR_{min}=100dB$}
  \end{axis}
\end{tikzpicture}
\caption{Error control performance of the ``JoComCo w prec" scheme.}
\label{pictureError}
\end{figure}
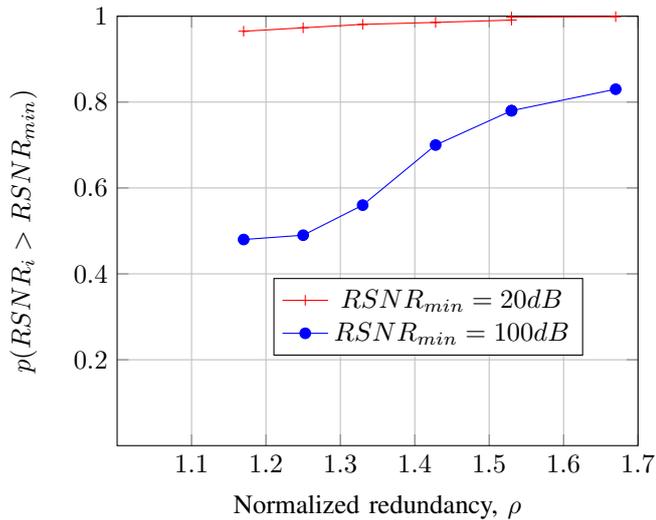

Figure \ref{pictureError} shows the reconstruction probability for two  RSNR bounds, namely $\text{RSNR}_{min}=20dB$ and $\text{RSNR}_{min}=100dB$. The number of packets including the coded ones is $l'=l\rho$ where $\rho$ is defined as the normalized redundancy. The packet loss probability in all the inter-cluster links  is set to $\epsilon=0.3$ and the packet redundancy added at each cluster head is varied around $\rho^*=\frac{1}{1-\epsilon}$. We remark that, there is still a good reconstruction probability even with lower redundancy than what is used for network coding scheme. This is a special feature of the spatial compression scheme and it is different from the all-or-nothing constraint of the classical random linear network coding. Intuitively, the reconstruction probability increases with higher redundancy. For $\text{RSNR}_{min}=20dB$ it is close to 1 and $\text{RSNR}_{min}=100dB$ it is above 0.7 for $\rho^*$.
\section{Conclusions and Perspectives}
This paper proposed a joint design of distributed compressed sensing and network coding with the underlying operations in the real field. An implementation of the proposed scheme for a specific cluster topology is carried out. Joint compressed sensing and random linear network coding has shown to be beneficial for exploiting the inherent correlation of the source data, error correction and opportunistic reception. We have also observed while implementing the scheme in \cite{feizi2011power} that the loss of RE property and the loss due to the quantization of real field operations can affect the performance of the algorithm in the topologies many hops. We have observed that it is optimal to use Bernoulli coefficients for intra-cluster network coding and Gaussian coefficients for inter-cluster network coding.  Furthermore, we proposed the optimal spatial pre-coding parameter $\mu$ and the number of network coded outgoing packets from each cluster head.

Knowing that compressed sensing can be designed in the finite fields \cite{draper2009compressed}, we would like to extend this topic into the finite fields in order to fully benefit from the features of network coding as a future direction.
\section*{Acknowledgement}
This work was financed partially by the CoSIP project
(FI 1671/1-1) granted by the German Research Foundation (DFG) and the Excellence Cluster Center for Advancing Electronics Dresden (CFAED).
\bibliographystyle{IEEEtran}
\bibliography{bibtex}
\end{document}